\documentclass[12pt,preprint]{aastex}
\usepackage{emulateapj5}
\usepackage{epsfig,natbib}
\slugcomment{To Appear in the Astrophysical Journal Supplements}
\shortauthors{Eracleous \& Hapern}
\shorttitle{Redshifts of Radio-Loud AGNs}

\def\apj{\rm {ApJ}}                
\def\apjs{\rm{ApJS}}               

\def\cd{c$\!\!\!\hskip 0.75pt$\raise 0.2pt \hbox{\symbol{24}}}

\def\Msol{\ifmmode{\rm M}_{\mathord\odot}\else M$_{\mathord\odot}$\fi}
\def\Mbh{\ifmmode{M_{\rm bh}}\else{$M_{\rm bh}$}\fi}

\def\ls{\lower 2pt \hbox{$\;\scriptscriptstyle \buildrel<\over\sim\;$}} 
\def\gs{\lower 2pt \hbox{$\;\scriptscriptstyle \buildrel>\over\sim\;$}}

\def\kms{\ifmmode{~{\rm km~s^{-1}}}\else{~km~s$^{-1}$}\fi}
\def\m#1{\ifmmode{^{-#1}}\else{$^{-#1}$}\fi}
\def\asec{\ifmmode{^{\prime\prime}}\else{$^{\prime\prime}$}\fi}
\def\asecb{\ifmmode{^{\prime\prime\!\!\!}}\else{$^{\prime\prime\!\!\!}$}\fi}
\def\asecp{\ifmmode{^{\prime\prime\!\!\!}.}\else{$^{\prime\prime\!\!\!}$.}\fi}
\def\deg{\ifmmode{^{\circ}}\else{$^{\circ}$}\fi}
\def\degp{\ifmmode{^{\circ\!\!\!}.}\else{$^{\circ\!\!\!}$.}\fi}

\def\ten#1{$10^{#1}$} 

\newcounter{species}
\def\ion#1#2{\setcounter{species}{#2}#1$\;${\sc\roman{species}}\relax}

\def\l{$\lambda$}

\def\a{$\alpha$}
\def\b{$\beta$}

\def\dzz{\delta z\sim}

\begin{document}

\title{Accurate Redshifts and Classifications for 110 Radio-Loud AGNs}

\author{Michael Eracleous\altaffilmark{1,2}}
\affil{Department of Astronomy and Astrophysics, The Pennsylvania State
University, 525 Davey Lab, University Park, PA 16803}
\and
\author{Jules P. Halpern\altaffilmark{1}}
\affil{Department of Astronomy, Columbia University, 550 West 120th St., 
New York, NY 10027}

\altaffiltext{1}{Visiting astronomer, Kitt Peak National Observatory,
which is operated by the AURA, Inc., under agreement with the National
Science Foundation} 

\altaffiltext{2}{Visiting astronomer, Cerro Tololo Inter-American
Observatory, which is operated by the AURA, Inc., under agreement with
the National Science Foundation}

\begin{abstract}
We report accurate redshifts of 110 active galaxies (mostly radio-loud
objects at $z<0.4$) observed in the course of a survey to find broad,
double-peaked emission lines. These redshifts are measured from the
narrow emission lines of these objects and are accurate to at least
one part in \ten{4}. For each object we determine a redshift from
high- and low-ionization lines separately, as well as an average
redshift from all the available lines. We find that in about 15\% of
cases, the low-ionization lines yield a slightly higher redshift than
the high-ionization lines; the average redshift difference amounts to
a velocity difference of approximately 80~\kms. In addition to the
redshift measurements we also report revised redshifts for two objects
as well as new classifications for three narrow-line objects.
\end{abstract}


\section{Introduction}

In the process of searching for broad, double-peaked Balmer emission
lines in moderate redshift, radio-loud active galactic nuclei (AGNs)
we surveyed 108 objects taken from AGN catalogs, {\it circa} 1991. A
detailed analysis of the results of this survey have been presented in
Eracleous \& Halpern (1994, 2003, hereafter papers I and II), where a
description of the observations and data reduction can also be found.
In this paper we report the redshifts measured from the spectra of
{\it all} objects observed in this survey, including spectra presented
in paper~I. The motivation for these redshift measurements is that
values accurate to $\dzz 10^{-4}$ are often needed for specific
applications (e.g., narrow-band imaging), while the cataloged values
have a precision of $\dzz 10^{-3}$. More importantly, we also find
that the values that we measure sometimes disagree with the cataloged
values at the level of $\dzz 10^{-3}$, and on some occasions at the
level of $\dzz 10^{-2}$.  Here we present redshifts, along with
estimated uncertainties which are of the order of $\dzz 10^{-4}$, and
often of the order of $\dzz\;{\rm a~few}\,\times 10^{-5}$.

In the process of acquiring the data we also observed 12 objects which
had only narrow emission lines and seven objects whose redshifts were
higher than 0.4. The spectra and revised redshifts of the first set of
such objects were presented in paper I. The second set of such objects
were observed after paper I was published and their spectra are
included in this paper for completeness (they were not included in
paper II so as to keep that paper short and focused).

\section{Redshift Measurements}

Our sample of objects consists of 89 objects from paper~I, 19 objects
from paper~II, plus Arp~102B and S4~0954+65 and includes broad-line
objects, as well as small number of narrow line objects (see paper~I
and \S2, below). The object S4~0954+65 is a blazar, which was included
in our sample unintentionally; we report its redshift here for
completeness. Its spectrum is shown by Lawrence et al. (1996).  The
spectra used for redshift determination were obtained during several
observing runs between 1988 and 2000 at the Kitt Peak National
Observatory's and Cerro Tololo Interamerican Observatory's 4m
telescopes, the Lick Observatory's 3m telescope, the MDM Observatory's
2.4m telescope and the Kitt Peak National Observatory's 2.1m
telescope.  We targeted the H$\alpha$ lines of objects in the
redshift range $z < 0.4$ thus, most of our spectra typically span a
wavelength range starting between 6,000 and 6,600~\AA\ and ending
between 9,500 and 10,000~\AA, with a resolution of 6--9~\AA.  For
objects with $z<0.1$, we obtained several spectra of the H$\alpha$ and
H$\beta$ region with starting wavelengths between 3,800 and 4,600~\AA\
and ending wavelengths between 7,400 and 8,500~\AA. In a few cases we
also observed objects with $0.1< z < 0.4$ at wavelengths as short as
3,200~\AA. As a result of the above observing strategies, the lines
most often available for redshift measurements were H$\alpha$ and the
low-ionization forbidden lines in its immediate vicinity (we were able
to measure H$\alpha$ in 92\% of objects, [\ion{N}{2}]~$\lambda$6583
and [\ion{O}{1}]~$\lambda$6300 in 2/3 of the objects each, and at
least one of the lines of the [\ion{S}{2}] doublet in approximately
half of the objects). The most commonly measured high-ionization lines
were those of the [\ion{O}{3}]~$\lambda\lambda$4959,5007 doublet
(available in about half of the objects, along with H$\beta$).
Finally in about 10--20\% of objects, we were able to measure lines in
the blue/near-UV part of the spectrum, namely \ion{He}{2},
[\ion{O}{3}]~$\lambda$4363, H$\gamma$, [\ion{Ne}{3}], [\ion{O}{2}],
and \ion{Mg}{2}.

Redshifts were determined by measuring the observed wavelengths of
reasonably strong {\it narrow} emission lines from all available
spectra. A list of the lines that have proven useful in this respect
is given in Table~\ref{Tlwave}, where we also identify which lines we
regard as high- and low-ionization lines. The rest wavelengths of
these lines were taken from Kaler et al. (1976).  In the case of the
[\ion{O}{2}], and \ion{Mg}{2} doublets, which are always unresolved in
our spectra, we adopted the average wavelength of the doublet as the
rest wavelength to be compared with our measurements.  The line
wavelengths were measured by fitting a Gaussian to several pixels
around the line peak (typically 7 to 11 pixels). Thus measurements
were made using only lines with clearly discernible peaks -- no
attempt was made to de-blend line complexes in which the line peaks
were not well separated. This technique allows us to locate the peak
of the line to better than a pixel, in particular to 0.30~pixels 50\%
of the time and 0.48~pixels 68\% of the time. After suitable
heliocentric velocity corrections the wavelengths of all lines from
the same object were used to determine its redshift. A redshift and
associated uncertainty were determined separately from low- and
high-ionization lines, and a mean redshift and associated uncertainty
were also determined using all available lines. The uncertainty in the
redshift was estimated as follows:

\begin{itemize}
\item
If two or more lines from the same object were available, the
uncertainty in the redshift was taken to be the adjusted error in the
mean, $s_{\rm n-1}$, given by (e.g., Barford 1985)

\begin{equation}
s^2_{\rm n-1} \equiv {\sigma^2\over{n-1}}=
{1\over{n(n-1)}} \sum^{n}_{i=1}(\bar z - z_{\rm i})^2 ,
\end{equation}

where $\bar z$ is the mean redshift, $n$ is the number of lines used,
and $z_{\rm i}$ is the redshift derived from an individual line
($\sigma$ is the usual root-mean-squared dispersion about the
mean). We note that $s_{\rm n-1}$ is related to the more usual error
in the mean, $s_{\rm n}$, by $s_{\rm n-1}=s_{\rm n}\;\sqrt{n/(n-1)}$,
where the factor $\sqrt{n/(n-1)}$ effectively takes into account the
fact that the sample has a finite membership (see Barford 1985 for a
more detailed discussion). This  statistical estimate of the
uncertainty accounts for several sources of error, such as errors in
locating the line peaks, and errors in the relative wavelength
calibration of a spectrum. If more than one spectrum is used, then
errors resulting from absolute differences in the wavelength scales of
individual spectra are also accounted for. However, this method does
not account for possible skewness or asymmetry of the line profiles,
and also assumes that all lines used in the redshift determination
have the same intrinsic redshift.

\item

If only a single emission line from a particular object was available,
the uncertainty in the redshift was estimated from the accuracy with
which the line peak can be located, namely the 68\% confidence limit
of 0.48 pixels quoted above. This limit was determined by considering
the dispersion about the mean redshift in objects where two or more
emission lines were available. As a result, it is not necessarily an
accurate estimate of the uncertainty in the redshift of a particular
object, but it does account for possible errors in the wavelength
calibration in an average sense, as well as errors in locating the
line peak.

\end{itemize}

The following sources of error deserve special mention and discussion.
All of these effects are captured by our statistical method for
estimating error bars.

\begin{enumerate}

\item
Systematic uncertainties associated with wavelength calibration are of
the order of a few $\times 10^{-5}$.

\item
The peaks of narrow lines that lie on the sloping wings of broad lines
could be slightly shifted. This is an issue with the Balmer and
[\ion{N}{2}] lines in most objects and also with the [\ion{S}{2}]
doublet, which lies on the red wind of the broad H$\alpha$ line in
about 15\% of our objects. To assess the magnitude of the effect, we
carried out tests in the most extreme cases of [\ion{S}{2}]
doublets. The tests consisted of comparing the peak wavelengths
measured with and without subtracting the underlying sloping
pseudo-continuum. We found that the difference in the results
translates into a median redshift difference of several $\times
10^{-5}$, which is comparable to the uncertainty in wavelength
calibration.

\item
The presence of \ion{Fe}{2} lines in the vicinity of the
[\ion{O}{3}]~$\lambda\lambda$4959,5007 doublet could affect the
measurement of the peaks of these lines. This could be important because
these lines are the most commonly measured high-ionization lines.  We
therefore inspected all of our spectra and did not find any cases with
discernible \ion{Fe}{2} lines in the regions of interest; thus we made
no attempt to remove them. The absence of \ion{Fe}{2} lines is
understandable since the vast majority of our objects are double-lobed
radio sources.

\end{enumerate}

The measured redshifts are given in Table~\ref{Tz}, which includes the
mean redshift (from all available lines), the low-ionization line
redshift and the high-ionization line redshift. Figures in parenthesis
give the uncertainty in the last digit of the reported redshift. For
reference the table also gives the number of spectra and the number of
low- and high-ionization lines used.

\section{Redshift Differences Between Low- and High-Ionization Lines}

To investigate whether the low- and high-ionization lines yield
significantly different redshifts, we have computed the difference
between these two redshifts for the 55 objects where both measurements
were available. The redshift difference was normalized by the its
uncertainty\footnote{The uncertainty in the redshift difference was
computed as $s_{\rm LH}=(s_{\rm LIL}^2+s_{\rm HIL}^2)^{1/2}$, where
$s_{\rm LIL}$ and $s_{\rm HIL}$ are the uncertainties in the redshifts
of the low- and high-ionization lines, respectively, given in
Table~\ref{Tz}}, to give $(z_{\rm LIL}-z_{\rm HIL})/s_{\rm LH}$, and
its distribution is shown in the histogram of Figure~\ref{Fzhist}.  If
the scatter in the redshift differences is entirely the result of
measurement errors, then we would expect the shape of the histogram to
resemble a Gaussian with unit standard deviation. Therefore, we have
overlaid in Figure~\ref{Fzhist} such a Gaussian, normalized so that
its mean square deviation from the bins of the histogram is minimized.
A visual comparison of this expected Gaussian distribution with the
observed histogram shows them to agree at negative values of the
redshift difference. At positive values, however, there appears to be
an excess of objects with $(z_{\rm LIL}-z_{\rm HIL})/s_{\rm LH} > 2$,
which suggests that in some cases the low-ionization lines yield a
larger redshift than the high-ionization lines. Isolating and
examining the objects that make up this excess, we find a mean
velocity difference between low- and high-ionization lines of
$c\langle z_{\rm LIL}-z_{\rm HIL}\rangle\approx 80\kms$. Although this
result is interesting, it needs to be verified using a much larger
sample of objects. If we accept it at face value, it may indicate a
stratification of the narrow-line region. More specifically, in the
context of a scenario in which the narrow-line region gas is
outflowing, this result implies that the high-ionization lines arise
in gas that is further away from the center and possibly of lower
density.

Redshift differences between high- and low-ionization lines in Seyfert
galaxies and narrow-line radio galaxies have been known for quite some
time. Koski (1978) compiled a list of 6 cases with redshift
differences of a few hundred \kms\ (600~\kms\ in the extreme case of
I~Zw~1), while Whittle (1985) and Veilleux (1991) found differences of
$\ls 100$\kms\ between the centroid of the [\ion{O}{3}]~\l5007 line
and the systemic velocity of the host galaxy. A more recent study of a
wide variety of AGNs by Marziani et al. (2003) yielded similar redshift
differences between the [\ion{O}{3}]~\l5007 and H\b\ lines and also
identified outliers with large redshift differences of order several
hundred \kms.  In the vast majority of cases the high-ionization lines
are blueshifted relative to the low-ionization lines, a trend which
supports the hypothesis that the narrow-line gas is outflowing from
the nucleus. The results of individual case studies of nearby
narrow-line AGNs that map the kinematics of the gas support this
hypothesis as well (e.g., Cygnus~A by Taylor, Tadhunter, \& Robinson
2003 and NGC~1068 by Cecil, Bland, \& Tully 1990).

\section{Narrow-Line Objects and New Redshifts}

\begin{figure}    
\begin{minipage}[t]{3.5in}
\centerline{\vbox to 3 truein{\vfill
\caption{The distribution of redshift differences between low- and
high-ionization lines normalized by its uncertainty. The histogram
includes 56 objects for which the necessary measurements were
available. The thick solid line is a Gaussian of unit standard
deviation, normalized so as to minimize its mean square deviation from
the bins of the histogram.
\label{Fzhist}}
\vfill
}}
\end{minipage}
\begin{minipage}[t]{3.5in}
\centerline{\plotone{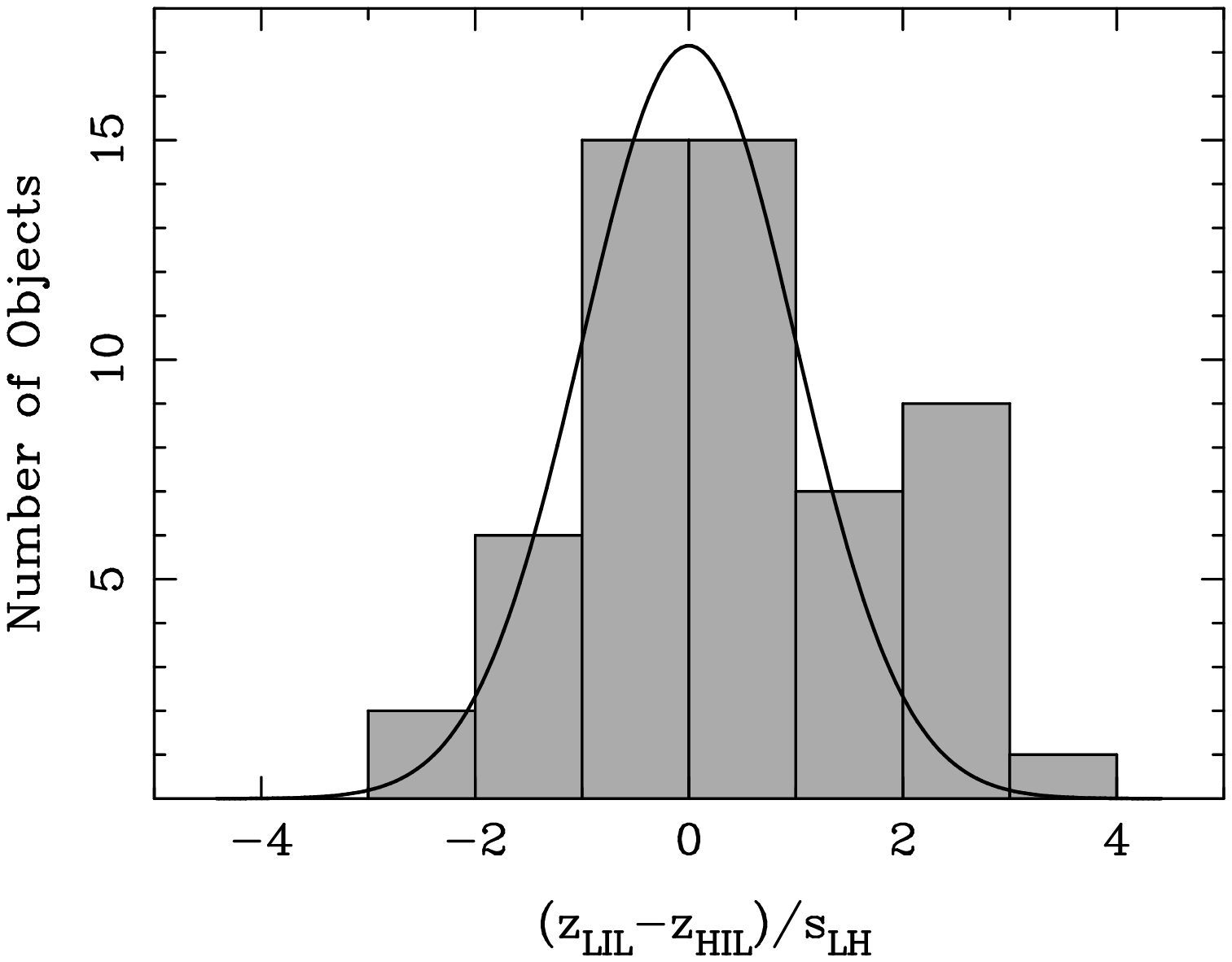}}
\end{minipage}
\begin{minipage}[t]{3.5in}
\centerline{\plotone{f2.eps}}
\caption{Spectra of three objects found to have only
narrow emission lines. These were reported as broad-line objects
in the literature.
\label{Fnarr}}
\end{minipage}
\hskip 0.3truein
\begin{minipage}[t]{3.5in}
\centerline{\plotone{f3.eps}}
\caption{ Spectra of two objects with incorrect redshifts reported in
the literature. The line identifications on which our redshift
determinations are based are marked, and the measured redshifts
are given in the figure.
\label{Fnewz}}
\end{minipage}
\end{figure}

Three objects listed in paper II, PKS~0511--48, 3C~381, and 3C~456,
were found to have only narrow emission lines; their spectra are shown
in Figure~\ref{Fnarr}. The radio galaxy 3C~381 was included in our
survey because of a report of a broad H\a\ line by Grandi \&
Osterbrock (1978). These authors noted that the signal-to-noise ratio
($S/N$) of the 3C~381 spectrum was low, but the spectrum was not shown
and the exact date of observation was not given (from indirect remarks
we infer that the observations were probably obtained in the early
1970s). A spectrum of 3C~381 obtained in 1982 by Saunders et
al. (1989) covering the rest-frame range 3100--6100~\AA\ with moderate
$S/N$ shows no broad lines. Our own spectrum shows this object as a
narrow-line radio galaxy. In view of the available data we conclude
that 3C~381 has either lost its broad lines some time in the past 20
years, or Grandi \& Osterbrock (1978) were misled in their detection
of a broad H\a\ line by the low $S/N$ of their spectrum.

Three objects listed in paper II, 4C~72.16, PKS~1355--12 and
PKS~2312--319, were found to have grossly incorrect cataloged
redshifts; as a consequence, their H\a\ lines did not fall within the
observed spectral range. These objects were included in our target
list because their redshifts listed in V\'eron-Cetty \& V\'eron (1989)
were lower than 0.4.  Since then 4C~72.16 was observed by Jackson \&
Browne (1991) who reported a redshift of 1.46, in agreement with our
own measurement; a plot of the spectrum can be found in that paper.
An incorrect redshift for 4C~72.16 of 0.357 was originally reported
by Arp, de Ruiter, \& Willis (1979), who noted that it was insecure.
PKS~1335--12 was observed by Stickel, Kuhr, \& Fried (1993)
who measured a redshift 0.539, in agreement with the value that we
obtain. The spectrum of this object presented in Figure~\ref{Fnewz}
covers a different spectral range than that of Stickel et
al. (1993). The redshift of PKS~2312--319 was reported to be 0.284 by
Jauncey et al. (1982), based on the identification of a single line at
3592~\AA\ as \ion{Mg}{2}~\l2800 (this line was observed in an
unpublished spectrum by R. G. Clowes, R. D. Cannon, and A. Savage
taken in 1980). We obtain a redshift of 1.322 based on the
identification of several emission lines. In view of this redshift,
the single emission line observed by Clowes et al. must have been
\ion{C}{4}~\l1550, which is also the strongest line in our own
spectrum. The spectrum of this object is shown in Figure~\ref{Fnewz}
where we also identify the lines that we used to determine its
redshift.

\acknowledgements We thank the anonymous referee for thoughtful
comments and suggestions. We acknowledge partial support from NASA
through grant GO-08684.01-A from the Space Telescope Science
Institute, which is operated by the Association of Universities for
Reasearch in Astronomy, Incorporated, under NASA contract NAS5-26555.



\clearpage
%
\begin{deluxetable}{lcc}
\tablewidth{0in}
\tablecolumns{2}
\tablecaption{Emission Lines Used for Redshift Determinations
and Adopted Rest Wavelengths\label{Tlwave}}
\tablehead{
\colhead{Ion and} &
\colhead{Adopted} &
\colhead{Low/High} \\
\colhead{Transition} &
\colhead{Wavelength (\AA)} &
\colhead{Ionization}
}
\startdata
\ion{Mg}{2}     $\lambda$2796,2803        & 2799.12 & L \\
{[\ion{O}{2}]}  $\lambda\lambda$3726,3729 & 3727.37 & L \\
{[\ion{Ne}{3}]} $\lambda$3869             & 3868.75 & H \\
                H$\gamma$                 & 4340.53 & L \\
{[\ion{O}{3}]}  $\lambda$4363             & 4363.21 & H \\
\ion{He}{2}     $\lambda$4686             & 4685.75 & H \\
                H$\beta$                  & 4861.30 & L \\
{[\ion{O}{3}]}  $\lambda$4959             & 4958.95 & H \\
{[\ion{O}{3}]}  $\lambda$5007             & 5006.88 & H \\
{[\ion{O}{1}]}  $\lambda$6300             & 6300.31 & L \\
{[\ion{O}{1}]}  $\lambda$6363             & 6363.82 & L \\
{[\ion{N}{2}]}  $\lambda$6548             & 6548.09 & L \\
                H$\alpha$                 & 6562.96 & L \\
{[\ion{N}{2}]}  $\lambda$6583             & 6583.36 & L \\
{[\ion{S}{2}]}  $\lambda$6717             & 6716.52 & L \\
{[\ion{S}{2}]}  $\lambda$6331             & 6730.74 & L \\
\enddata
\end{deluxetable}
\clearpage
\begin{deluxetable}{llllrrrc}
\tablewidth{0in}
\tablecolumns{8}
\tablecaption{Measured Redshifts\tablenotemark{a} \label{Tz}}
\tablehead{
\colhead{\hbox to 8em{\hfil Object\hfil}} &
\colhead{\hbox to 5em{\hfil $\bar z$\hfil}} &
\colhead{\hbox to 5em{\hfil $z_{_{\rm LIL}}$\hfil}} &
\colhead{\hbox to 5em{\hfil $z_{_{\rm HIL}}$\hfil}} &
\colhead{\hbox to 3em{\hfil Spectra\hfil}} &
\colhead{\hbox to 3em{\hfil LILs}} &
\colhead{\hbox to 3em{\hfil HILs}} &
\colhead{\hbox to 3em{\hfil Notes\tablenotemark{b}}} 
}
\startdata
4C 25.01            & 0.2844(2)   & 0.2844(2)   &  \dots       &  1 &  3 &    &  \\
3C 17               & 0.22024(7)  & 0.22019(7)  &  0.2205(1)   &  2 & 15 &  3 &  \\
MRC 0041+119        & 0.2261(3)   & 0.2263(3)   &  0.2258(5)   &  1 &  3 &  3 &  \\
PKS 0105--00\tablenotemark{c}   & 1.375(1)    & 1.375(1)    &  \dots       &  1 &  1 &    & new $z$ \\
B2 0110+29          & 0.3627(2)   & 0.3627(2)   &  \dots       &  1 &  3 &    &  \\
PKS 0152--51\tablenotemark{c}   & 1.582(1)    & 1.582(1)    &  \dots       &  1 &  1 &    & new $z$ \\
4C 31.06            & 0.37334(8)  & 0.37342(9)  &  0.3732(1)   &  1 &  5 &  3 &  \\
PKS 0202--76        & 0.38925(4)  & 0.38930(2)  &  0.38919(6)  &  1 &  3 &  2 &  \\
3C 59               & 0.1096(1)   & 0.1096(1)   &  \dots       &  2 & 10 &    &  \\
PKS 0214+10         & 0.4070(1)   & 0.4070(2)   &  0.40709(2)  &  1 &  2 &  2 &  \\
3C 67               & 0.3107(2)   & 0.3107(2)   &  \dots       &  1 &  3 &    &  \\
PKS 0235+023        & 0.2072(1)   & 0.2072(1)   &  \dots       &  2 &  9 &    &  \\
IRAS 0236.6--3101   & 0.06233(3)  & 0.06233(3)  &  \dots       &  1 &  5 &    &  \\
4C 39.11            & 0.16099(3)  & 0.16099(3)  &  \dots       &  1 &  5 &    &  \\
PKS 0312--77        & 0.2250(5)   & 0.2250(5)   &  \dots       &  1 &  3 &    &  \\
PKS 0340--37        & 0.28513(4)  & 0.28509(5)  &  0.28526(3)  &  3 & 15 &  5 &  \\
3C 93               & 0.35730(6)  & 0.35728(9)  &  0.35735(1)  &  2 &  6 &  2 &  \\
3C 93.1             & 0.24425(9)  & 0.24425(9)  &  \dots       &  1 &  5 &    & NLRG \\
3C 111              & 0.04907(6)  & 0.04909(9)  &  0.04902(2)  &  1 &  5 &  2 &  \\
3C 119              & 1.0221(3)   & 1.0221(7)   &  1.02207(1)  &  1 &  2 &  1 & new $z$ \\
MS 0450.3--1817     & 0.06161(7)  & 0.06161(7)  &  \dots       &  2 & 11 &    &  \\
PKS 0511--48        & 0.30638(4)  & 0.30636(6)  &  0.30641(2)  &  1 &  4 &  2 & NLRG \\
3C 135              & 0.1273(1)   & 0.1273(1)   &  \dots       &  1 &  6 &    & NLRG \\
Pictor A            & 0.03498(5)  & 0.03495(5)  &  0.0350(1)   &  1 &  6 &  3 &  \\
PKS 0558--50        & 0.1379(2)   & 0.1379(2)   &  \dots       &  1 &  1 &    &  \\
3C 171              & 0.2382(1)   & 0.2382(1)   &  \dots       &  1 &  5 &    & NLRG \\
PKS 0723--008       & 0.1273(1)   & 0.1273(1)   &  \dots       &  1 &  6 &    & NLRG \\
B2 0742+31          & 0.46134(1)  & 0.46134(2)  &  0.46134(2)  &  3 &  4 &  6 &  \\
PKS 0812+02         & 0.4028(2)   & 0.4028(2)   &  \dots       &  1 &  2 &    &  \\
CBS 74              & 0.09190(2)  & 0.09188(2)  &  0.091974(6) &  2 & 17 &  5 &  \\
3C 206              & 0.1973(1)   & 0.1973(1)   &  \dots       &  1 &  5 &    &  \\
DW 0839+187         & 1.2724(4)   & 1.2724(4)   &  \dots       &  1 &  1 &    & new $z$ \\
PKS 0846+10         & 0.36546(9)  & 0.36546(9)  &  \dots       &  1 &  2 &    &  \\
PKS 0857--19        & 0.3608(1)   & 0.3608(1)   &  0.36081(8)  &  1 &  5 &  2 &  \\
4C 5.38             & 0.30157(3)  & 0.30160(3)  &  0.30151(2)  &  1 &  5 &  2 &  \\
PKS 0921--213       & 0.05308(2)  & 0.05306(2)  &  0.05313(5)  &  2 & 12 &  6 &  \\
PKS 0925--203\tablenotemark{d}  & 0.34735(9)  & 0.3475(1)   &  0.34723(2)  &  1 &  2 &  2 &  \\
3C 227              & 0.08603(5)  & 0.08608(6)  &  0.0859(1)   &  4 & 14 &  7 &  \\
4C 9.35             & 0.29809(2)  & 0.29809(3)  &  0.29808(2)  &  1 &  5 &  2 &  \\
S4 0954+65          & 0.3675(2)   & 0.3675(2)   &  \dots       &  1 &  3 &    & blazar \\
PKS 1004--21\tablenotemark{d}   & 0.33046(7)  & 0.33053(1)  &  0.3303(4)   &  1 &  2 &  1 &  \\
PKS 1004+13         & 0.2406(1)   & 0.2406(1)   &  \dots       &  1 &  4 &    &  \\
PKS 1011--282       & 0.2549(1)   & 0.2549(1)   &  \dots       &  1 &  4 &    &  \\
PKS 1020--103       & 0.1965(1)   & 0.1965(1)   &  \dots       &  3 & 10 &    &  \\
B2 1028+31          & 0.17785(9)  & 0.17785(9)  &  \dots       &  1 &  5 &    &  \\
3C 246              & 0.34545(7)  & 0.3455(2)   &  0.345421(2) &  1 &  2 &  2 &  \\
PG 1049--006        & 0.3585(6)   & 0.3585(6)   &  \dots       &  1 &  5 &    &  \\
4C 72.16\tablenotemark{c}       & 1.462(1)    & 1.462(1)    &  \dots       &  1 &  1 &    & new $z$ \\
PKS 1101--32        & 0.35547(5)  & 0.35553(7)  &  0.35538(2)  &  1 &  3 &  2 &  \\
4C 36.18            & 0.39220(3)  & 0.39219(5)  &  0.39224(2)  &  2 & 10 &  5 &  \\
B2 1128+31          & 0.2897(1)   & 0.2899(2)   &  0.28954(2)  &  1 &  1 &  2 &  \\
PKS 1146--037       & 0.34056(8)  & 0.34056(8)  &  \dots       &  1 &  6 &    &  \\
LB 2136             & 0.33345(4)  & 0.33345(4)  &  \dots       &  1 &  4 &    &  \\
PKS 1151--34        & 0.25799(6)  & 0.25801(7)  &  0.2579(1)   &  4 & 23 &  7 &  \\
TEX 1156+21         & 0.3472(2)   & 0.3472(2)   &  \dots       &  1 &  1 &    &  \\
B2 1208+32A         & 0.38879(9)  & 0.38879(9)  &  \dots       &  1 &  2 &    &  \\
PKS 1215+013        & 0.11715(1)  & 0.11715(1)  &  \dots       &  1 &  5 &    & NLRG \\
B2 1223+25          & 0.2679(1)   & 0.2678(2)   &  0.2680(3)   &  1 &  3 &  1 &  \\
PKS 1232--24        & 0.35512(6)  & 0.35507(9)  &  0.35522(1)  &  1 &  4 &  2 &  \\
3C 277.1            & 0.3198(2)   & 0.3198(2)   &  \dots       &  1 &  5 &    &  \\
PKS 1254--33        & 0.19048(9)  & 0.19048(9)  &  \dots       &  1 &  2 &    &  \\
B2 1255+37          & 0.7092(2)   & 0.7090(3)   &  0.7093(3)   &  1 &  1 &  2 & new $z$ \\
PKS 1302--102       & 0.27825(8)  & 0.27825(8)  &  \dots       &  1 &  2 &    &  \\
PKS 1304--215       & 0.12666(9)  & 0.12666(9)  &  \dots       &  1 &  5 &    & NLRG \\
3C 287.1            & 0.21574(5)  & 0.21574(5)  &  \dots       &  1 &  6 &    &  \\
PKS 1335--12        & 0.5386(1)   & 0.5384(4)   &  0.53869(7)  &  1 &  1 &  2 & new $z$ \\
PKS 1346--11        & 0.3407(1)   & 0.3408(1)   &  0.34050(4)  &  1 &  4 &  2 &  \\
B2 1351+26          & 0.30763(7)  & 0.30763(7)  &  \dots       &  1 &  3 &    &  \\
PKS 1355--41        & 0.3145(1)   & 0.3145(2)   &  0.314657(6) &  1 &  3 &  2 &  \\
Mkn 668             & 0.07681(7)  & 0.07688(8)  &  0.07665(6)  &  4 &  9 &  4 &  \\
PKS 1417--19        & 0.12039(9)  & 0.12039(9)  &  \dots       &  1 &  5 &    &  \\
PKS 1421--38        & 0.4068(2)   & 0.4068(3)   &  0.40687(1)  &  1 &  3 &  2 &  \\
CSO 643             & 0.27615(3)  & 0.27615(3)  &  0.27616(3)  &  2 & 10 &  4 &  \\
3C 303              & 0.14121(5)  & 0.14121(5)  &  \dots       &  1 &  7 &    &  \\
PKS 1451--37        & 0.3143(2)   & 0.3143(3)   &  0.31428(2)  &  1 &  6 &  2 &  \\
4C 37.43            & 0.37086(7)  & 0.3708(1)   &  0.37092(6)  &  1 &  3 &  3 &  \\
PKS 1514+00         & 0.0526(2)   & 0.0528(2)   &  0.0521(2)   &  1 &  6 &  2 &  \\
LB 9743             & 0.2536(1)   & 0.2536(1)   &  \dots       &  1 &  4 &    &  \\
4C 35.37            & 0.1565(2)   & 0.1565(2)   &  \dots       &  1 &  5 &    &  \\
4C 18.47            & 0.34647(7)  & 0.34647(7)  &  \dots       &  1 &  3 &    &  \\
3C 332              & 0.15098(2)  & 0.15099(2)  &  0.150910(9) &  3 & 16 &  2 &  \\
MRC 1635+119        & 0.1474(2)   & 0.1474(2)   &  \dots       &  1 &  3 &    &  \\
3C 351              & 0.37194(4)  & 0.37189(6)  &  0.37200(4)  &  2 &  4 &  3 &  \\
Arp 102B            & 0.02436(3)  & 0.02437(3)  &  0.02436(6)  &  3 & 27 & 12 &  \\
B2 1719+35          & 0.28345(3)  & 0.28344(4)  &  0.28348(2)  &  1 &  6 &  2 &  \\
B2 1721+34          & 0.2053(1)   & 0.2053(1)   &  \dots       &  1 &  4 &    &  \\
PKS 1725+044        & 0.29662(8)  & 0.29662(8)  &  \dots       &  1 &  4 &    &  \\
PKS 1739+18C        & 0.18588(6)  & 0.18595(8)  &  0.18577(2)  &  1 &  3 &  2 &  \\
MRC 1745+16         & 0.3919(1)   & 0.3919(1)   &  0.39185(2)  &  1 &  6 &  2 &  \\
3C 381              & 0.16058(4)  & 0.16061(6)  &  0.16052(5)  &  2 & 10 &  7 & NLRG \\
3C 382              & 0.05807(5)  & 0.05806(6)  &  0.058156(8) &  4 & 17 &  2 &  \\
3C 390.3            & 0.05550(2)  & 0.05551(2)  &  0.05548(3)  &  6 & 29 &  3 &  \\
PKS 1914--45        & 0.36378(3)  & 0.36376(5)  &  0.36381(1)  &  1 &  4 &  2 &  \\
PKS 2058--42        & 0.2232(1)   & 0.2232(1)   &  \dots       &  1 &  4 &    &  \\
PKS 2139--04        & 0.34390(9)  & 0.3440(1)   &  0.343780(5) &  1 &  3 &  2 &  \\
OX 169\tablenotemark{e}  & 0.21087(5)  & 0.21114(3)  &  0.21061(6)  & 12 & 19 & 20 &  \\
PKS 2159--335       & 0.1537(3)   & 0.1537(3)   &  \dots       &  1 &  2 &    & NLRG \\
PKS 2208--13\tablenotemark{d}   & 0.39110(8)  & 0.3912(2)   &  0.39104(1)  &  1 &  2 &  2 &  \\
3C 445              & 0.05623(2)  & 0.05623(2)  &  0.05622(5)  &  3 & 13 &  4 &  \\
PKS 2227--399       & 0.31788(7)  & 0.3179(1)   &  0.317971(8) &  1 &  6 &  2 &  \\
PKS 2242--29        & 0.1658(2)   & 0.1658(2)   &  \dots       &  1 &  2 &    & NLRG \\
PKS 2247+14         & 0.23478(8)  & 0.2349(1)   &  0.23466(5)  &  1 &  5 &  3 &  \\
PKS 2300--18        & 0.12883(6)  & 0.12883(6)  &  \dots       &  1 &  6 &    &  \\
PKS 2302--71        & 0.38425(6)  & 0.3842(1)   &  0.38429(2)  &  1 &  2 &  2 &  \\
PKS 2305+18         & 0.31301(4)  & 0.31305(4)  &  0.31297(7)  &  1 &  7 &  5 &  \\
3C 456              & 0.2326(2)   & 0.2328(2)   &  0.2323(4)   &  1 &  6 &  4 & NLRG \\
PKS 2312--319\tablenotemark{f}  & 1.322(1)    & 1.324(2)    &  1.321(1)    &  1 &  2 &  3 & new $z$ \\
3C 459              & 0.2201(2)   & 0.2201(2)   &  \dots       &  1 &  4 &    & NLRG \\
MRC 2328+167\tablenotemark{d}   & 0.2801(5)   & 0.2795(3)   &  0.2806(9)   &  1 &  3 &  3 &  \\
PKS 2349--01        & 0.1741(1)   & 0.1741(1)   &  \dots       &  1 &  4 &    &  \\
\enddata
\tablenotetext{a}{
$\bar z$, $z_{_{\rm LIL}}$, and $z_{_{\rm HIL}}$ denote respectively
the redshifts derived from all available lines, low-ionization lines only, and 
high-ionization lines only. The figure in parenthesis is the uncertainty in the last
digit of the reported redshift.}
\tablenotetext{b}{
``NLRG'' = narrow-line object, ``new $z$'' = revised redshift (from paper~I
or paper~II).}
\tablenotetext{c}{
The redshift is based on the measured wavelength of the \ion{Mg}{2} doublet only,
whose peak is often broad and asymmetric, and also affected by associated
absorption lines.  Therefore the uncertainty is larger than usual.}
\tablenotetext{d}{
The only available low-ionization lines are the
Balmer lines whose peaks are rather broad. This suggests that the narrow Balmer
lines are extremely weak and  hence the measured LIL redshift refers to the
broad lines.}
\tablenotetext{e}{The redshift of OX~169 was measured using all the spectra 
from Halpern \& Eracleous (2000)}
\tablenotetext{f}{
The lines used for the determining the redshift of PKS~2312--319 are:
\ion{C}{4}~$\lambda$1549, \ion{N}{3}]~$\lambda$1750,
\ion{C}{3}]~$\lambda$1909, \ion{Mg}{2}~$\lambda$2798, and
[\ion{O}{2}]~$\lambda$3727. The signal-to-noise ratio is low and
therefore the uncertainty in the redshift is larger than usual.}
\end{deluxetable}

%
\end{document}